\documentclass[12pt,preprint]{aastex}

\slugcomment{ApJ Letters accepted}

\shorttitle{Blue HB stars in the Sgr dSph}
\shortauthors{Monaco et al.}
\def\gtsima
{\hbox{\raise0.5ex\hbox{$>\lower1.06ex\hbox{$\kern-1.07em{\sim}$}$}}}
\def\ltsima
{\hbox{\raise0.5ex\hbox{$<\lower1.06ex\hbox{$\kern-1.07em{\sim}$}$}}}
\begin{document}

\title{Blue Horizontal Branch Stars in the \\
Sagittarius dwarf spheroidal galaxy}

\author{Lorenzo Monaco} \affil{Dip. di Astronomia, Universit\`a di
Bologna, Via Ranzani 1, 40127,  Bologna, ITALY \\ INAF - Osservatorio
Astronomico di Bologna, Via Ranzani 1, 40127, Bologna, 
ITALY;\email{lorenzo@uqbar.bo.astro.it}} 

\author{Michele Bellazzini} \affil{INAF - Osservatorio Astronomico di
Bologna, Via Ranzani 1, 40127, Bologna,  ITALY}

\author{Francesco R. Ferraro} \affil{Dip. di Astronomia, Universit\`a
di Bologna, Via Ranzani 1, 40127,  Bologna, ITALY}

\and

\author{Elena Pancino} \affil{INAF - Osservatorio Astronomico di
Bologna, Via Ranzani 1, 40127, Bologna, ITALY}

\email{bellazzini,ferraro,pancino@bo.astro.it}


\begin{abstract}  

We report on the recovery of a Blue Horizontal Branch
(BHB) population belonging to the Sagittarius Dwarf Spheroidal Galaxy
(Sgr). The sequence is clearly identified in the (V, V-I) Color
Magnitude Diagram (CMD) obtained for about 500,000 stars in the region of 
the globular cluster M~54. The BHB
morphology is similar to the analogous sequence in M~54, but it is
unambiguously associated with Sgr since {\it(i)} it is detected well
outside the main body of the cluster, up to more than 5 tidal radii from the cluster center and
{\it(ii)} the BHB stars follow the radial distribution of the other stellar
populations of Sgr. This finding finally demonstrates that the Sgr
galaxy hosts a significant (of the order of $\sim$10\%) old and
metal-poor stellar population ([Fe/H]\ltsima -1.3; age \gtsima 10 Gyr),
similar to that of its oldest clusters (M~54, Ter~8). We also show that
the Sgr BHB sequence found here is the counterpart of the analogous 
feature observed by Newberg et
al. (2002) in the Sgr Stream, in a field more than 80$\degr$ away from
the center of the galaxy.

\end{abstract}



\section{Introduction}

It is generally recognized that the Sagittarius dwarf Spheroidal galaxy
\cite[][]{s1,s2} is currently disrupting under the strain of the Milky
Way tidal field. The stars \cite[and clusters, see][]{mb1,mb2} lost by
the Sgr dSph remain coherently aligned along the orbital path of the
galaxy, forming a huge filamentary structure, the Sgr Stream, that has
now been observed over the entire sky \cite[][ and references
therein]{carb,orb-sds,orb-2mass,yanny,ivez,sdss,david,kundu,mb2,steve}. 
Therefore, the identification and definition of all the stellar
components belonging to the Sgr dSph is crucial not only to understand
the Star Formation History (SFH) and chemical evolution of this
keystone galaxy, but also to trace its remnants in the Galactic Halo.

The Color Magnitude Diagram (CMD) of the Sgr galaxy is dominated by the
classical features of a metal-rich population, with a red and extended
Red Giant Branch (RGB) and a Red Clump (RC) of He-burning stars 
\cite[][]{musk,sl95,marc97,sdgs1,sdgs2,bump}. It is now
generally recognized that this dominant population has a mean age of
4-6 Gyr (\citet[][]{ls00} hereafter LS00, \citet[][]{bump}, hereafter
Pap-I) and a high mean metallicity \citep[][ LS00, Pap-I]{tammy,bc03}.

The apparent absence of old and metal poor stars (OMS) in Sgr was noted
since the early studies and persists, in part, at the present
day. 
However, OMS should be present, since (a) previous generations of
stars are required to enrich the medium from which the dominant
metal-rich population has formed and (b) the Sgr dSph hosts at least
two classical old and metal poor globular cluster \citep[Ter~8
and M~54, see][ and LS00, respectively]{mont}. Hence, we have observational
evidence that OMS were formed in Sgr, even if just in globular
clusters. Unfortunately, the search for RGB stars bluer than the main
RGB of Sgr (hence, presumably, more metal poor) has been hampered by
the strong contamination by foreground stars affecting the CMD.
Therefore, the available spectroscopic surveys included only a
negligible fraction of stars with $[Fe/H]<-1$ \citep{tammy,bc03}. 

A Blue Horizontal Branch (BHB) sequence is easily recognizeable in 
the wide-field photographic CMD of \citet{igi} and it can be considered a 
clear indication of the presence 
of OMS in Sgr: however this population has never been further observed in any 
successive Sgr CCD survey.
Some additional indirect clues of the existence of a metal-poor
population in Sgr exist in the literature, in particular: 

\begin{enumerate} 

\item \citet{sdgs1} surveyed an isolated field in Sgr and found a star 
count excess in a region of the CMD similar in magnitude and color to the HB
of Ter~8. However, due to the few (23) Sgr BHB candidates, this was considered
just as a supporting evidence that old and metal poor stars exist in the field
of Sgr.

\item \citet{cser} analized the light curve of $\sim$1800 RR Lyrae in the 
Sgr galaxy. From the location of the RRd variables in the Petersen diagram he
concluded that RR Lyrae in Sgr have metallicities in the range 
$-2.0$\ltsima[Fe/H]\ltsima$-1.3$ \cite[see also LS00;][]{sam}. Since RR
Lyrae stars are thought to trace old populations (\gtsima 10 Gyr), this
result is a convincing proof that some old and metal poor 
stars are present in
the Sgr field population. According to \citet{cser}, the central
density of RR Lyrae in Sgr is $\simeq 139$ stars/deg$^2$. Since we
found (Pap-I) $\sim$2000 stars/deg$^2$ belonging to the Red Clump, the RR
Lyrae should account for just $\sim$6\% of the whole HB population.
\end{enumerate}
In this letter we use the CMD of a large area ($1 \times 1$ deg$^2$) in
the main body of the Sgr dSph (Pap-I) to show that a
significant BHB population is indeed present in Sgr. We also show that
the BHB is most likely the counterpart of the analogous feature found
in the Sgr Stream by \citet[][]{sdss}.

\section{The Blue HB of Sgr}

The database presented in Pap-I contains V and I photometry of
$\sim$490,000 sources down to $V\sim 23$ in a $1 \times 1$ deg$^2$ field
centered on the globular cluster M~54, which coincides with the
position of the maximum density peak of the Sgr galaxy (see Pap-I for
further details). 

Figure 1 shows a zoom of the HB region of the CMD, at different
distances from the center of M~54. Panel (a) refers to stars contained
within the tidal radius of M~54\footnote{In the following, $r_t$ must
always be intended as the tidal radius of the globular cluster M~54.}
\cite[$r<r_t$, where $r_t =7\arcmin.5$, ][]{trag}. As can be seen,
although the galactic foreground stars (vertical structure at
V-I$\simeq$0.8) and the main Sgr population  (Red Clump at $V\simeq
18.2$) are easily recognizable, the population of M~54 obviously
dominates the CMD. In particular the cluster RGB, ranging from (V,
V-I)$\sim$(21, 0.9) up to (16.4, 1.3), and the well known BHB, at
V-I\ltsima0.4 from $V\simeq 18$ down to $V\sim 21$, are clearly evident
(see also LS00 and references therein). The BHB sequence of M~54 has
been enclosed in a box that is repeated in all panels.

Panel (b) shows the CMD of stars lying at r$>$r$_t$: as expected, 
while the RGB of M~54 disappears, the
galactic foreground and the Sgr population dominate the CMD at this
distance from M~54. However, the most surprising feature is the
unexpected  presence of a conspicuous population of BHB (see also panels
(c) and (d)).  The BHB population (and in particular its brightest bulk
at  V$<$18.6) remains quite numerous even in the most external part 
($3r_t \leq r \leq 6r_t$) of the surveyed region (see panel (d)).

Figure 1 thus demonstrates that a significant BHB population is indeed
present in the main body of the Sgr galaxy, as previously
suggested by \citet{igi} and \citet{sdgs1}. The morphology of the BHB found in Sgr
is quite similar to that of M~54, although the faintest part of the
sequence (V$>$18.6) appears less populated with respect to M~54.

In order to further test the connection between the BHB population and
the Sgr galaxy, we have selected two different stellar samples (shown
in the left panel of Figure 2) and we have compared their radial distributions. The first
sample contains stars belonging to the RC and
therefore traces the distribution of stars belonging exclusively to the
Sgr galaxy. The second contains stars belonging to the brightest part
of the BHB only (V$<$18.6) to avoid spurious effects due to
incompleteness and contamination by the Blue Plume. The
BHB sample should therefore trace the BHB of M~54 plus the 
contribution of the Sgr BHB. 

Figure 3 shows the cumulative radial distribution of the two adopted
test samples outside $r_t$ (left panel) and $2r_t$ (right panel). 
As can ben seen from both panels of Figure 3 the two samples appear remarkably 
similar. A Kolmogorov-Smirnov test provides a large probability ($>$66\%) 
that the two distributions are extracted from the same parent population, 
supporting 
the idea that BHB stars outside M~54 actually belong to the Sgr galaxy.

Figure 4 shows instead the cumulative distribution of the two 
samples for $r>3\arcmin$, well inside the tidal radius of M~54, but
still far enough from the very core of the cluster, where crowding
becomes severe. The cumulative radial profile of M~54 is 
represented by a King model \citep{king} with $r_c=0\arcmin.11$ and
c=1.84 \citep{trag}. From the inspection of the figure 
it is then evident that, even inside $r_t$, the
cumulative distribution of the BHB sample is significantly different from
the star distribution in M~54. In fact, to reproduce the observed
BHB sample distribution, we need to use a combined population. 
Figure 4 shows that the global distribution of BHB stars is nicely reproduced 
by a composite population with (approximately) $\sim$30\% of stars contributed 
by M~54 and $\sim$70\% by Sgr. However, a good agreement can also be
achieved by varying the relative fractions by $\sim$10\%.  

\section{Discussion} 

In the previous section we report convincing evidence that a
substantial population of BHB is present in the main body of Sgr. The
existence of this population was missed by all previous photometric CCD
surveys, probably due to an unsufficient field of view. This recovery
can shed light on different aspects of the structure and the
history of this disrupting stellar system.

\subsection{BHB: tracing the Old and Metal poor stars in Sgr} 

The presence of a well-populated BHB in Sgr implies the presence of a
substantial population of OMS in this galaxy. In fact, BHB have been, up
to now, observed only in old, metal-poor stellar populations. The only exceptions are two metal-rich and very dense globular
clusters belonging to the bulge \citep{rich}. However, BHB stars appear
spatially segregated toward the center of these clusters, suggesting a
dynamical origin for these stars \citep{lay}. Due to the extremely low stellar density
conditions, this kind of formation channel appears quite unlikely in
the case of the Sgr galaxy. Thus, with a morphology that is
very similar to that of the old, metal-poor clusters associated to the
Sgr: Ter~8 ([Fe/H]$=-1.99$) and M~54 ([Fe/H]$=-1.55$), the BHB sequence of Sgr is more
likely composed of genuine OMS stars. 

Moreover, although it is often difficult to place precise constraints
on age and metallicity from the study of the HB morphology
alone\footnote{We recall here that the HB morphology depends on a {\it
combination of parameters}, including age and metallicity, but also
less understood processes like the mass loss during the RGB phase.}, we
can anyway put some limits to the age and metallicity of the BHB
population by observing that stars bluer than the RR Lyrae should be at
least as old and metal-poor as the RR Lyrae themselves. In
particular RR Lyrae stars are known to trace old populations 
($\gtsima$10 Gyr). Hence, by combining this with the results by \citet{cser}, 
it can be concluded that {\it the Sgr BHB population is composed
by old and metal-poor stars, with ages greater than $\sim$10 Gyr and
metallicities lower than [Fe/H]$\simeq$-1.3}.

We can also estimate which fraction of the HB stars is composed by OMS,
by comparing the stellar densities of RC stars, RR Lyrae and BHB stars.

We select all the RC stars with $r>r_t$ (lower selection box 
in the right panel of Figure 2).
Then we estimate the level of foreground contamination by counting
stars within a box adjacent to the RC sample box (upper 
selection box in the right panel of Figure 2)
having the same size, and we obtain N$_f$$\sim$1000 stars/0.95~deg$^2$ 
(we consider all stars with $r>r_t$, i.e. a total area of 0.95~deg$^2$). 
We use this estimate to
clean the RC sample, obtaining a final RC density of N$_{RC}$$\sim$2000
stars/0.95~deg$^2$. In the same region, we counted stars belonging to
the bright BHB sample defined in Figure 2, obtaining  N$_{BHB}=$147
stars/0.95~deg$^2$. Finally, from the work of \citet{cser}, 
the expected number of RR Lyrae is N$_{RR}$$\simeq$130 stars/0.95~deg$^2$.

We conclude that {\it the fraction of old and metal-poor
stars in the Sgr galaxy is $\sim$12\%.} 
Therefore, while the most important episodes of star formation in the Sgr 
galaxy occurred between 4 and 6 Gyr ago (Pap-I, 
corresponding\footnote{A cosmology with $h=0.65$, $\Omega_{matter}=0.3$ and
$\Omega_{\Lambda}=0.7$ has been assumed.} to
$z<1$) from a quite metal-rich ([M/H]\gtsima -0.6)
interstellar medium, a non negligible fraction of metal poor stars were 
already in place more than 
10 Gyr ago ($z$\gtsima2).

\citet{sdgs1} estimated a larger, but still compatible, 
fraction of OMS: $\leq$30\%.  
However it has to be noted that our estimate should be considered a lower limit 
since a larger fraction can be obtained if fainter BHB are taken into account.

\subsection{BHB stars in the Sgr Stream}
 
Recently \citet{sdss}, using the Early Data Release of the Sloan
Digital Sky Survey, SDSS\footnote{See http://www.sdss.org.}, report on the
discovery of an unexplained stellar population in a region of the sky
named S341+57-22.5, in the direction $(l,b)\simeq (341,+57)$. The
authors suggest that the unknown population is part of the Sgr Stream
since (1) S341+57-22.5 lies on the predicted Sgr's orbital path
\cite[according to][]{orb,orb-sds} and (2) the CMD is quite similar to
that of the Sgr galaxy published by \citet{marc97}, with the notable exception
of a BHB sequence which was lacking in any previous CMD of
Sgr \citep[with the only exception of the photographic CMD by][]{igi}.

We used the same equations that \citet{sdss} applied to the (V,V-I)
photometry of \citet{marc97} to transform our photometry into the SDSS
photometric system, adopting the same distance and reddening as well. 
The BHB presented in this Letter is then located in the CMD 
at $-0.3< g^*-r^* < 0.0$, a position which is fully compatible with the
BHB sequence discovered by \citet{sdss}. Thus, the BHB population found
by \citet{sdss} in the Sgr stream corresponds to the BHB population
found in the Sgr galaxy. This implies that the disruption of the Sgr
dSph provided a sizeable contribution to the assembly of the Galactic
Halo not only in the form of metal-rich stars \cite[traced by the M
giants studied by][]{orb-2mass,mb2,steve} but also providing a
substantial amount of old and metal-poor stars that constitute the
typical halo population.

\acknowledgments

We acknowledge the financial support from the italian
Ministero dell'Universit\`a e della Ricerca (MIUR) and from the Agenzia Spaziale
Italiana (ASI).


\begin{figure}
\plotone{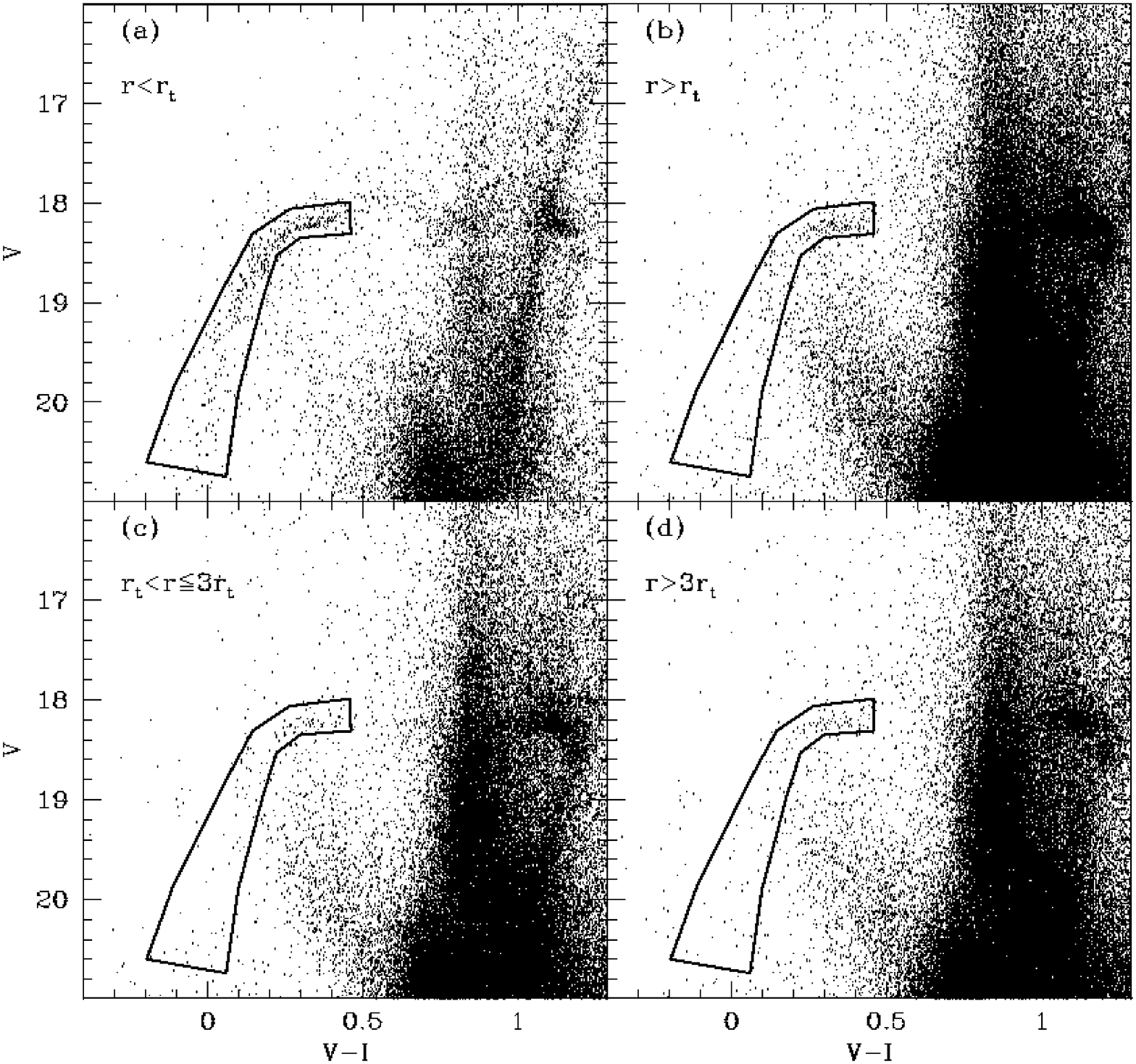}
\caption{CMDs at different radial distances from the center of the
globular cluster M~54. The thick line encloses stars belonging to the
BHB in all panels.}
\end{figure}

\begin{figure}
\plotone{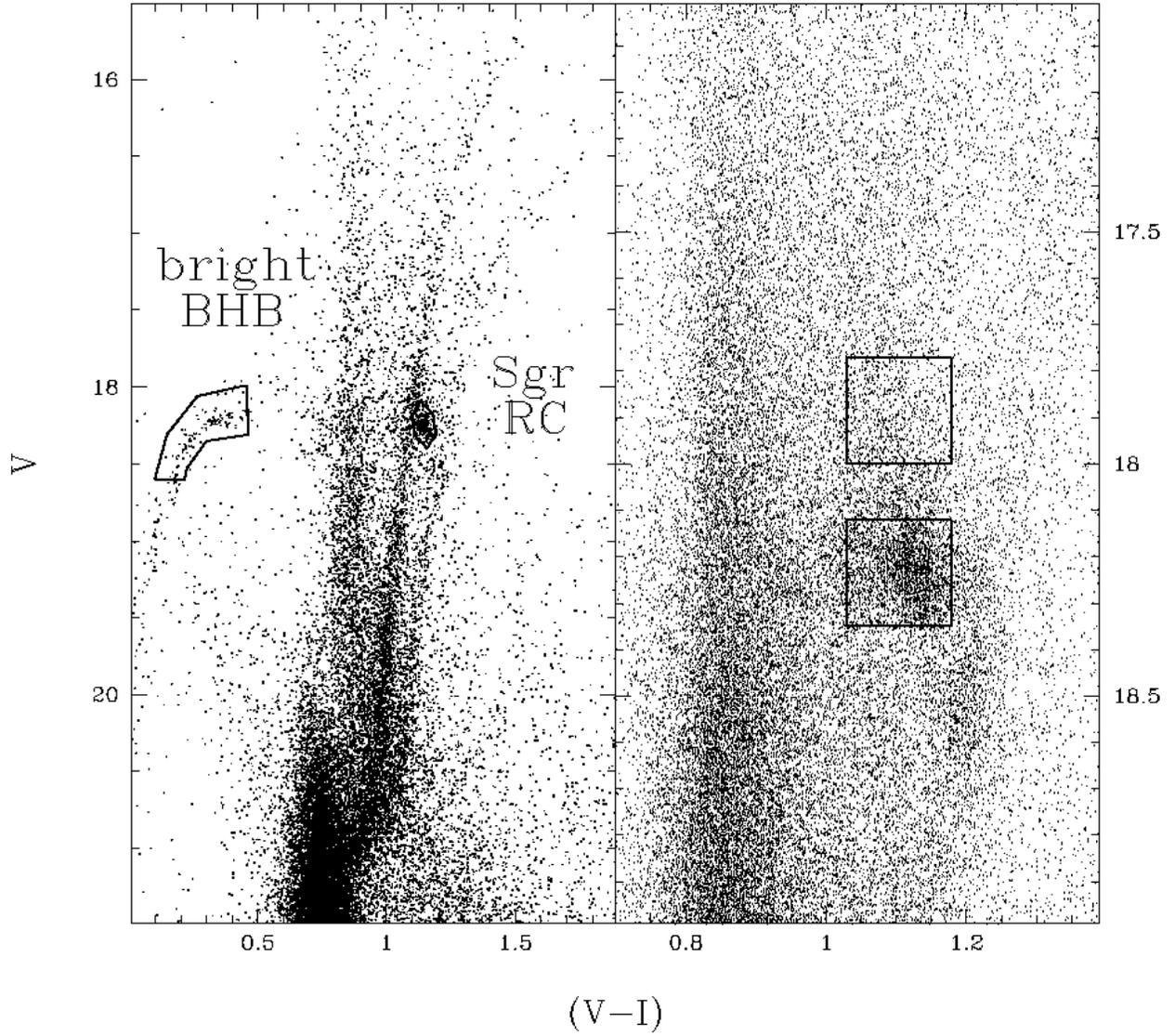}
\caption{Left panel: adopted selection boxes for BHB and RC stars as tracer
populations on the 
Sgr CMD. Right panel: CMD of Sgr outside the tidal radius of M~54. The lower 
box encloses stars in the RC phase, while the 
upper selection box is used for field decontamination.}
\end{figure}

\begin{figure}
\plotone{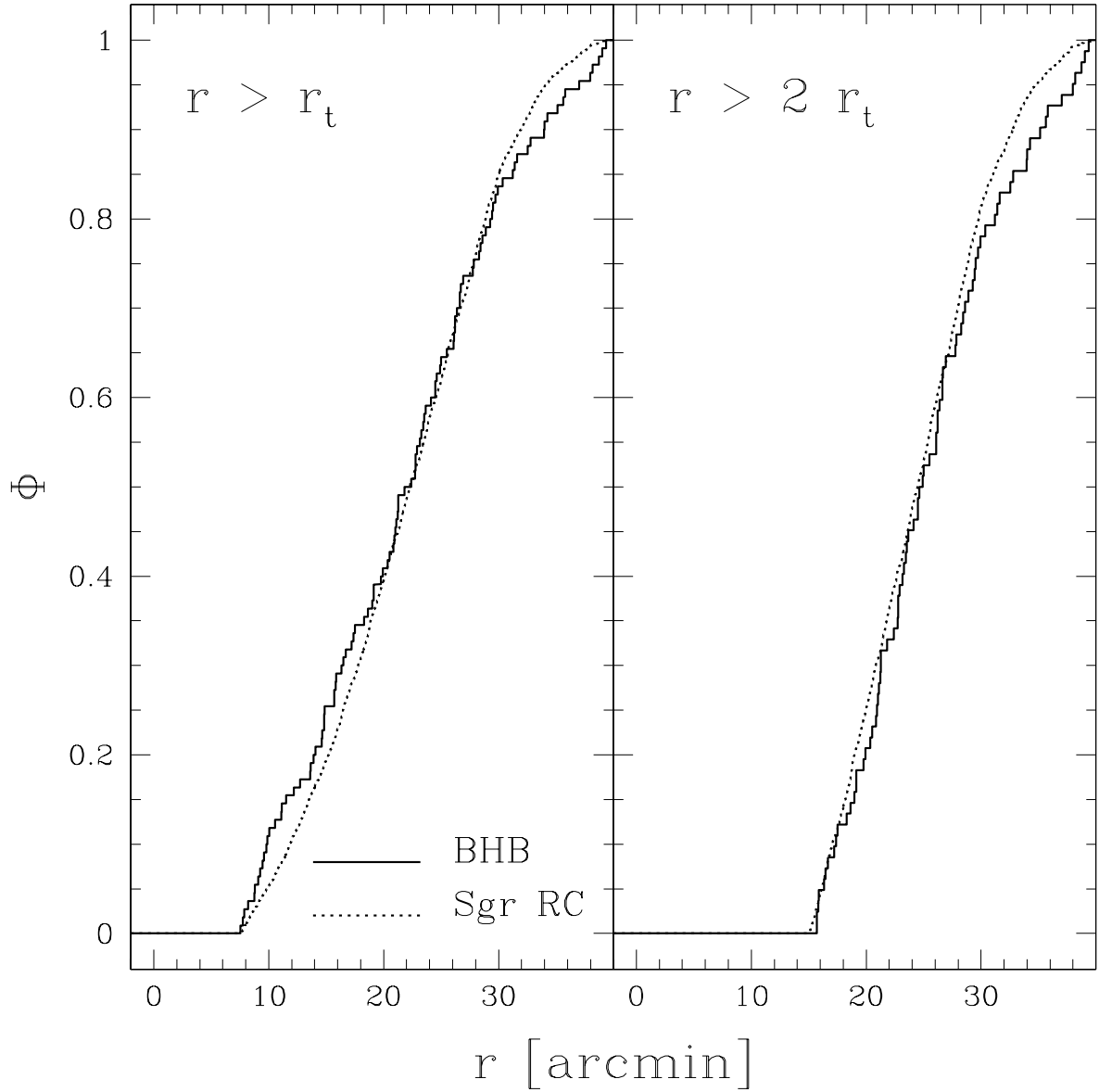}
\caption{The cumulative radial distribution of the BHB sample (solid
line) is compared with that of the RC control sample (dotted line),
outside $r_t$ (left panel) and $2r_t$ (right panel) from M~54.}
\end{figure}

\begin{figure}
\plotone{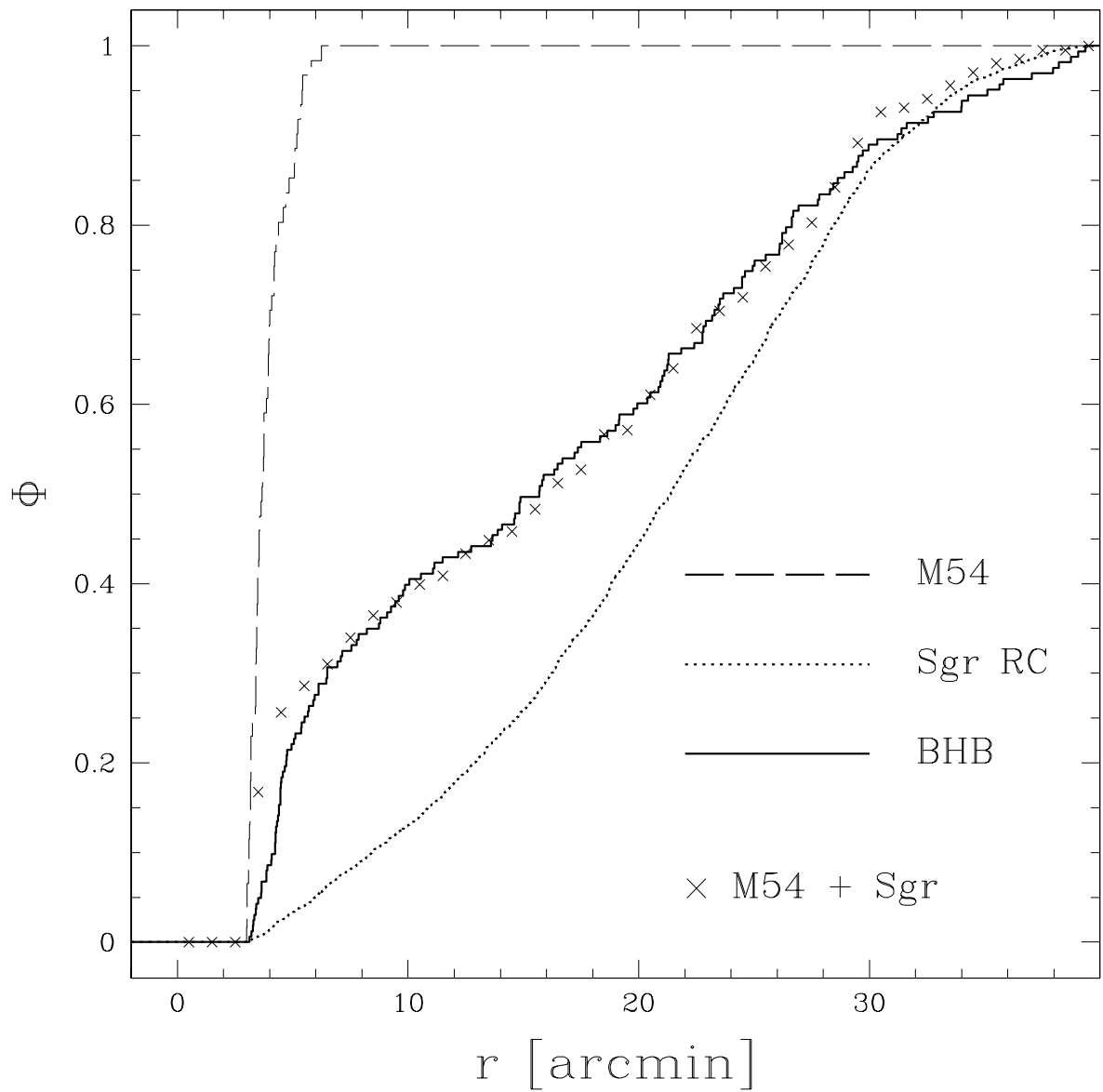}
\caption{The cumulative radial distribution of the BHB sample (solid
line), starting at $3 \arcmin$ from the center of M~54, is compared
with the distribution of the RC sample (dotted line) and of M~54
(dashed line), represented by its King profile \citep{king} with
$r_c=0\arcmin.11$ and c=1.84 \citep{trag}. Crosses represent a suitable
combination of the profile of M~54 ($\sim$30\%) and that of the
Sgr dSph ($\sim$70\%).}

\end{figure}

\end{document}